# Quantum simulation of honeycomb lattice model by high-order moiré pattern

Qiang Wan[1§], Chunlong Wu[1§], Xun-Jiang Luo[2§], Shenghao Dai[1], Cao Peng[1], Renzhe Li[1], Shangkun Mo[1], Keming Zhao[1], Wen-Xuan Qiu[2], Hao Zhong[1], Yiwei Li[1], Chendong Zhang[2,3], Fengcheng Wu[2,3*], and Nan Xu[1,3*]

[1] *Institute of Advanced Studies, Wuhan University, Wuhan 430072, China*
[2] *School of Physics and Technology, Wuhan University, Wuhan 430072, China*
[3] *Wuhan Institute of Quantum Technology, Wuhan 430206, China*

§ There authors contributed equally.
* E-mail: nxu@whu.edu.cn; wufcheng@whu.edu.cn

**Moiré superlattices have become an emergent solid-state platform for simulating quantum lattice models. However, in single moiré device, Hamiltonians parameters like lattice constant, hopping and interaction terms can hardly be manipulated, limiting the controllability and accessibility of moiré quantum simulator. Here, by combining angle-resolved photoemission spectroscopy and theoretical analysis, we demonstrate that high-order moiré patterns in graphene-monolayered xenon/krypton heterostructures can simulate honeycomb model in mesoscale, with *in-situ* tunable Hamiltonians parameters. The length scale of simulated lattice constant can be tuned by annealing processes, which *in-situ* adjusts intervalley interaction and hopping parameters in the simulated honeycomb lattice. The sign of the lattice constant can be switched by choosing xenon or krypton monolayer deposited on graphene, which controls sublattice degree of freedom and valley arrangment of Dirac fermions. Our work establishes a novel path for experimentally simulating the honeycomb model with tunable parameters by high-order moiré patterns.**

# I. INTRODUCTION

Moiré superlattices provide not only a rich playground for the exploration of novel quantum phases [1-6], but also a new quantum simulator of solid state lattice models [7-9]. The flat bands in magic-angle twisted bilayer graphene possess a fragile topology, which obstructs deriving a simple tight-binding model [10-11]. However, recent theoretical progress revealed that twisted bilayer graphene effectively realizes a topological heavy fermion model with coexisting localized and itinerant electrons [12]. Moiré superlattices formed by transition metal dichalcogenides (TMD) were theoretically proposed to simulate quantum models on triangular lattices [7-8] or honeycomb lattices [13-14], depending on the exact material combinations. Experimental discoveries of Mott insulators [15-16], generalized Wigner crystal states [17-18], and topological states [19] in TMD-based moiré superlattices strongly support the feasibility of quantum simulation.

Although electron filling can be continuously tuned by electric gating, Hamiltonian parameters like lattice constant, single-particle hopping, and many-body interaction strength cannot be easily manipulated in moiré devices formed between two layers with similar lattice constants and small twist angles [20-24], limiting the controllability and accessibility of moiré quantum simulator.

Recently, high-order moiré patterns were realized in graphene-based heterostructures with large lattice constant mismatch $\Delta a$, including graphene on SiC and graphene-monolayered xenon heterostructure (mXe/G) [25-28]. Strikingly, the high-order moiré pattern in mXe/G can be *in situ* tuned by controlling the lattice constant of the xenon monolayer through annealing, which further generates movable Dirac cone replicas in the electronic structure as observed by angle-resolved photoemission spectroscopy (ARPES) [26]. An open question is whether high-order moiré pattern can provide a new route to realize a quantum simulator of model Hamiltonians.

Here, by combining ARPES experiment and theoretical study, we unveil that the high-order moiré pattern in graphene-noble gas monolayer heterostructure can be a good experimental platform to simulate mesoscale honeycomb lattice, with not only tunable magnitude but also a switchable sign of lattice constant. The magnitude of the honeycomb lattice constant, equivalent to the moiré period $a'_m$, can be *in situ* tuned by annealing processes, which determines the hopping term and intervalley interaction strength in the simulated mesoscale honeycomb lattice. By choosing Xe or Kr monolayer deposited on graphene, the sign of $a'_m$ can also be changed with opposite sublattice arrangement, which is directly evidenced by the different locations of moiré Brillouin zone (BZ) near the K point and the opposite valley arrangement of the Dirac cones in mXe/G and graphene-Kr monolayer heterostructure (mKr/G) as observed in ARPES measurements. These experimental observations are consistent with our theoretical analysis, which indicates different sublattice arrangements for the two systems. By varying annealing temperature, the magnitude of $a'_m$ can be *in situ* tuned, which controls the intervalley coupling strength and the hopping parameters in the tight-binding model on the honeycomb lattice. Our work demonstrates that the high-order moiré pattern in mXe/G and mKr/G is a promising experimental platform to simulate a mesoscale honeycomb model with *in situ* tunable parameters.

## II. EXPERIMENTAL DETAILS

Monolayer graphene was grown by a high-temperature annealing process [29] of n-type 6H-SiC(0001) from PrMat. Kr/Xe monolayers were grown on monolayer graphene at T = 30 K with Kr/Xe pressure of $1\times10^{-9}$ Torr, for a duration of one hour to deposit rare gases. Subsequently, the samples were gradually heated at a rate of 0.2 K/min to reach the corresponding annealing temperatures. The ARPES measurements were performed at lab-based ARPES facility utilizing He lamp as the photon source (He Iα: 21.2 eV). We use the same graphene sample and measurement geometry in this work to avoid the difference from effect of matrix element for the direct comparison between results of mXe/G and mKr/G.

## III. RESULTS

Firstly, we compare the high-order moiré patterns in mXe/G and mKr/G, as sketched in Fig. 1. The lattice constants of Xe monolayer ($a_{Xe}$) and Kr monolayer ($a_{Kr}$) are slightly larger [Fig. 1(b)-I] and smaller [Fig. 1(e)-I] than that of graphene $\sqrt{3} \times \sqrt{3}$ supercell $\sqrt{3}a_{Gra}$, respectively. The moiré period of mXe/G and mKr/G can be defined in a unified manner as

$$a'_m = \frac{a_{Gra}a_{Xe(Kr)}}{a_{Xe(Kr)} - \sqrt{3}a_{Gra}} \quad (1).$$

With this definition, there is a sign difference in the moiré period $a'_m$ between mXe/G and mKr/G, which leads to different stacking configurations in mXe/G and mKr/G. The distinction is particularly pronounced at $A_M$ and $B_M$ sites, which are two inequivalent high-symmetry positions in the moiré pattern. At the $A_M$ sites, the Xe atom overlaps with the B-sublattice of graphene in mXe/G [Fig. 1(b)-II], but the Kr atom overlaps with the A-sublattice in mKr/G [Fig. 1(e)-II]. On the other hand, at the $B_M$ sites, the Xe atom overlaps with the A-sublattice of graphene in mXe/G [Fig. 1(b)-III], but the Kr atom overlaps with the B-sublattice in mKr/G [Fig. 1(e)-III]. The stacking patterns between graphene and noble gas monolayers are experimentally deteremined by Low Energy Electron Diffraction (LEED) measurements (Fig. S1 in the Supplemental Material [30]).

The sign difference of $a'_m$ also has a significant impact on the electronic structure modulation in mXe/G and mKr/G. In mXe/G, the Dirac replicas in the moiré BZ have the same valley arrangement as that of graphene [Fig. 1(c)], as the reciprocal lattice vector of Xe monolayer is less in magnitude than the Γ-K vector of graphene ($|\vec{G}_{Xe}| \leq |\vec{K}_{Gra}|$). For the negative $a'_m$ moiré pattern in mKr/G, the Dirac replicas have an opposite valley arrangement [Fig. 1(f)] compared to that of both bare graphene and mXe/G, because of $|\vec{G}_{Kr}| \geq |\vec{K}_{Gra}|$. For the same reason, the replicas near the K point locate outside the graphene first BZ in mKr/G [Fig. 1(f)], in contrast to the case of mXe/G [Fig. 1(c)].

The effect of $a'_m$ sign difference is directly observed by ARPES measurements, as both mXe/G and mKr/G possess an identical graphene layer and maintain a consistent geometric relationship with the helium lamp. Figure 1(g)-(h) present the Fermi surface (FS) maps of mXe/G and mKr/G, respectively, with $a_{Xe} \sim 4.50$ Å and $a_{kr} \sim 4.04$ Å, exhibiting similar $|a'_m| \sim 4.600$ nm. The most pronounced difference is that Dirac cone replicas near the K point sit inside the graphene BZ in mXe/G, but appear outside the graphene BZ in mKr/G, fully consistent with the discussion in Fig. 1(c) and 1(f).

The different valley arrangement induced by the $a'_m$ sign difference in mXe/G and mKr/G is also demonstrated by the ARPES results on the Dirac cone replicas near the Γ point. Away from the Dirac point, the original graphene Dirac cone shows a valley-contrast trigonal warping. Therefore, the different valley arrangements of Dirac cone replicas in the moiré BZ in mXe/G and mKr/G induce obvious differences in the dispersions at high $E_B$, as clearly seen from constant energy plots in ARPES experiments [Fig. 1(i) and 1(j)]. For mXe/G, the constant energy plots show a flower-like pattern at high $E_B$ in Fig. 1(i). In contrast, mKr/G shows a wheel-like pattern for constant energy plots with the same $E_B$ in Fig. 1(j). The differences are consistent with the different valley arrangement of the warping Dirac cones as indicated by guiding dashed lines in Fig. 1(i) and 1(j).

The moiré pattern period in mXe/G and mKr/G can be *in situ* tuned, which covers a wide range in both positive and negative directions. As pointed out by our previous work [26], the moiré BZ in mXe/G shrinks by decreasing the annealing temperature $T_a$ [Fig. 2(a)], indicating an increase of moiré period $a'_m$ in mesoscale. The mechanism is that the lattice constant of the Xe monolayer $a_{Xe}$ becomes smaller and approaches the value of $\sqrt{3} a_{Gra}$ by decreasing $T_a$. According to Eq. (1), the moiré period $a'_m$ is positive and increases with lowering $T_a$. Correspondingly, the moiré BZ shrinks [Fig. 2(a)] and the Dirac replicas located at the opposite moiré BZ corners $\kappa_m$ and $\kappa_m'$ move closer to each other in mXe/G at lower $T_a$, as clearly presented by the band dispersion evolution along the momentum path #1 [Fig. 2(b)]. When $a_{Xe}$ is equal to $\sqrt{3}a_{Gra}$ at $T_a$ = 51.5 K, the moiré pattern evolves into the commensurate Kekulé distortion [Fig. 2(a)-IV and 2(b)-IV].

On the other hand, in the case of mKr/G, the Kr monolayer also has smaller lattice constants $a_{Kr}$ with decreasing $T_a$. However, in the range of $a_{Kr} \leq \sqrt{3}a_{Gra}$, the smaller $a_{Kr}$ results in a negative $a_m'$ with a smaller absolute value $|a_m'|$, according to Eq. (1). Correspondingly, the Dirac replicas at $\kappa_m$ and $\kappa_m'$ move away from each other in mKr/G with lowering $T_a$ [Fig. 2(b)], resulting in an expanding moiré BZ with the opposite valley arrangements compared to that in mXe/G and original graphene [Fig. 2(a)].

The distinction between moiré patterns in mXe/G and mKr/G can be better visulized in $T_a$ dependent photoemission data at the Γ point. While the doulbe peaks become close to each other with increasing $T_a$ in mXe/G [Fig. 2(c)], mKr/G shows the opposite behavior [Fig. 2(d)]. For Kekulé distortion as Dirac replicas overlap with each other, the spectra weight at the Dirac point is suppressed and a double-peaks line shape with an energy separation of 0.48 eV is observed at the Dirac point in both mXe/G and mKr/G, in contrast to the single-peak profile in original Dirac cone observed in printine graphene [Fig. 2(e)]. Although the gap size at Dirac point can be overestimated in ARPES measurements [28], our results clearly moiré modulation effect on Dirac band structure in mXe/G and mKr/G.

Moiré period $a_m'$ can also be tuned by different annealing pressure of noble gas. By increasing the Kr pressure from $1.33 \times 10^{-9}$ torr to $6.65 \times 10^{-8}$ torr at $T_a$ = 46 K [Fig. 2(f)], mKr/G evolves from Kekulé distortion ($|a_m'| \to \infty$) to $|a_m'|$ = 4.81 nm [Fig.

2(g)]. The moiré pattern with a specific value $a'_m$ can be achived at high temperature, by increasing Kr pressure. Therefore, ultra-low temperature and ultra-high vacuum conditions are in fact not necessary for achieving moiré pattern in mXe/G and mKr/G.

After the annealing processes, we can keep the moiré pattern at low temperature by cooling the system in an environment without noble gas partial pressure. At the lowest temperature we can achieve in our measurements (T = 6 K), we manage to achieve moiré pattern with $|a'_m|$ from 4.32 nm to ∞ by different annealing temperatures [Fig. 2(h)-(i)]. Therefore, this route also provides a feasible way for the low-temperature transport and spectral measurement of mKr/G to resolve the possible flat moiré band and related interaction-driven quantum phases.

The results in Fig. 2 indicate that, by tunning annealing temperature and pressure, the high-order moiré periodicity in mXe/G and mKr/G complementarily covers almost the full range of $a'_m$ in both positive and negative directions, as summarized in Fig. 2(j).

To gain an in-depth understanding of the experimental results, we perform theoretical analysis, which indicates that the high-order moiré pattern in mXe/G and mKr/G are good platforms to simulate the honeycomb lattice with tunable sublattice degree of freedom. We theoretically construct the low-energy continuum moiré Hamiltonians for mXe/G and mKr/G,

$$H = \begin{pmatrix} h_+(\mathbf{k} - s\mathbf{\kappa}_+) & T_s(\mathbf{r}) \\ T_s^\dagger(\mathbf{r}) & h_-(\mathbf{k} - s\mathbf{\kappa}_-) \end{pmatrix},$$

where the diagonal terms $h_\pm(\mathbf{k}) = \hbar v_F(\pm k_x \sigma_x + k_y \sigma_y)$ are the Dirac Hamiltonians for the two valleys of monolayer graphene with $\sigma_{x,y}$ being Pauli matrices in the A and B sublattice space, the index $s$ is $+1$ ($-1$) for mXe/G (mKr/G), and $\mathbf{\kappa}_\pm = 4\pi/(3|a'_m|)(\mp 1/2, \sqrt{3}/2)$ are located at the corners of the moiré Brillouin zone. The off-diagonal term $T_s(r)$ is the intervalley coupling induced by the moiré superlattices and is parametrized as follows:

$$T_s(r) = T_0 + T_{+1}e^{isg_2\cdot r} + T_{-1}e^{isg_3\cdot r},$$

$$T_j = \begin{pmatrix} w_0 e^{-i2\pi j/3} & w_1 \\ w_1 & w_0 e^{i2\pi j/3} \end{pmatrix},$$

where $g_2 = 4\pi/(\sqrt{3}|a'_m|)(-\sqrt{3}/2, 1/2)$ and $g_3 = 4\pi/(\sqrt{3}|a'_m|)(-\sqrt{3}/2, -1/2)$ are moiré reciprocal lattice vectors. By fitting to Kekulé gap of the experiment data, we have $w_1 = 67$ meV for mXe/G and $w_1 = 73$ meV for mKr/G, respectively. By fitting to first-principles band structures (see Supplemental Material [30]), we estimated that $w_0 = 1.33$ meV for mXe/G and $w_0 = 0.58$ meV for mKr/G, respectively. Here $w_0$ is an order of magnitude smaller than $w_1$.

We display the theoretical moiré band structure of mKr/G with $a'_m = 5$ nm in Fig. 3(a). Our model clearly indicates two middle moiré bands connected by the Dirac points at the charge neutrality point. The symmetrized ARPES band structure near the K point of graphene [black line in Fig. 3(b)-I)] are shown in Fig. 3(b), with $a'_m = 5$ nm, 10 nm and 20 nm. To directly compare the experimental results, the theoretical ARPES results based on the calculated band structure are also appended. The sudden intensity changes in ARPES results support the hybridization gaps at the crossing points between the main Dirac cone and replicas as indicated by the simulations, and reveal the formation of moiré bands.

Symmetry analysis of these two moiré bands indicates that they can be effectively described by a tight-binding model defined on the honeycomb lattice formed by $A_M$ and $B_M$ sites [Fig. 3(c)], with the lattice constant equivalent to $a'_m$. We construct the two Wannier states centered, respectively, at $A_M$ and $B_M$ sites. Using the obtained Wannier states, we further calculate the hopping parameters in the tight-binding model, which accurately reproduces the energy dispersion of the two middle bands [Fig. 3(a)]. For mXe/G, a similar conclusion can also be achieved. Namely, the two middle moiré bands in mXe/G also match with the results of the tight-binding model on the honeycomb lattice formed by $A_M$ and $B_M$ sites. However, there is a crucial difference regarding the sublattice degree of freedom between mXe/G and mKr/G. The Wannier state centered at $A_M$ ($B_M$) site is polarized to A(B) sublattice in mXe/G, but B(A) sublattice in mKr/G. This difference is revealed by the distinct real-space patterns of sublattice polarization for the two middle moiré bands in mXe/G and mKr/G, as shown in Fig. 3(d)-(e). We note that the different sublattice patterns in real space are counterparts of different valley arrangements in momentum space for the two systems, which could be experimentally detected by atomic scanning tunneling spectroscopy. In Fig. 3(f), we present the envlution of $\left|t^{(1)}_{A_M B_M}\right|$ (the nearest-neighbor hopping parameter) as a function of $a'_m$, which decreases rapidly accounting for the decreasing of bandwidth with the increasing of $a'_m$.

For completeness, we finally present the topological phase diagram of the moiré Hamiltonians as a function of model parameters. Up to an energy constant, the Hamiltonian $H$ is fully determined by two dimensionless parameters $w/\hbar v_F |\kappa_+|$ and $\beta$, where $w$ is equal to $\sqrt{w_0^2 + w_1^2}$ and $\beta$ is defined through $(w_0, w_1) = w(\sin\beta, \cos\beta)$. In Fig. 4, we present the topological phase diagram for the two middle moiré bands as a function of $w/\hbar v_F |\kappa_+|$ and $\beta$. The color shaded regions represent gapped phases, where the two middle bands are separated by an energy gap from other bands. Based on symmetry eigenvalues and Wilson loop spectra, the gapped phases can be further classified into fragile topological phases protected by $\hat{C}_{2z}\hat{T}$ symmetry and topologically trivial phases described by a honeycomb lattice model. Here $\hat{C}_{2z}$ is the twofold rotation symmetry around the z axis and $\hat{T}$ is time-reversal symmetry. In the limit of $w_1 = 0$ and $\beta = \frac{\pi}{2}$, there is an exact mapping of $H_{\text{Xe/Kr}}$ to the chiral limit of twisted bilayer graphene [31-32], which is consistent with the fact that the fragile topological phases appear around $\beta = \pi/2$ in our phase diagram. Our systems mXe/G and mKr/G are in the opposite limit of $w_0 \ll w_1$, where the two middle bands are topologically trivial and belong to phase (4) in Fig. 4. We note that parameters $w_0$ and $w_1$ are system dependent; the phase diagram in Fig. 4 shows the possibility of realizing fragile topological phases in other graphene superlattice systems.

## IV. CONCLUSIONS

In summary, the *in situ* tunable high-order moiré patterns in mXe/G and mKr/G provide good platforms to simulate honeycomb lattices in mesoscale. As summarized in Fig. 2(j), by tuning the temperature and partial pressure in the annealing process, the honeycomb lattice constant $a'_m$ in mXe/G and mKr/G covers the range in mesoscale along positive and negative directions, respectively. In the meantime, the hopping term in the tight-binding model between the nearest-neighbor $A_M$ and $B_M$ decreases rapidly with increasing $a'_m$ [Fig. 3(f)]. This is expected since the bandwidth of the two middle bands decreases when $a'_m$ increases, as indicated in Fig. 3(b). A honeycomb lattice model with narrow or nearly flat bands can host strong many-body correlation effects and will serve as a testbed for exotic phases that are highly sought-after theoretically [33-35]. Our work will stimulate further theoretical and experimental studies to reveal the novel properties of mesoscale graphene.

**APPENDIX**

**Excluding photoelectron diffraction effect**

The observation of replica bands can not be explained by photoelectron diffraction effect [36]. In the photoelectron diffraction scenario as illustrated in Fig. 5(a), photoelectrons from graphene Dirac cone are diffracted by Kr monolayer, resulting in replica band with the same position as $\kappa'_m$ [Fig. 5(b)]. However, the original photoelectrons have a limitation of momentum parallel to the sample surface, $K_\parallel = \frac{\sqrt{2m_e E_{kin}}}{\hbar}$, where $E_{Kin}$ and $m_e$ are photoelectrons' kinetic energy and mass. Correspondingly, the diffraction induced replica bands are also expected to have a cutoff [Fig. 5(a)]. In Fig. 5(b), we clearly observe that $\kappa'_m$ replica exceed the limitation, which contradicts to photoelectron diffraction scenario. Therefore, our results confirm that the observed replicas are induced by moiré potential modulated initial state, rather than photoelectron diffraction.

# Figures

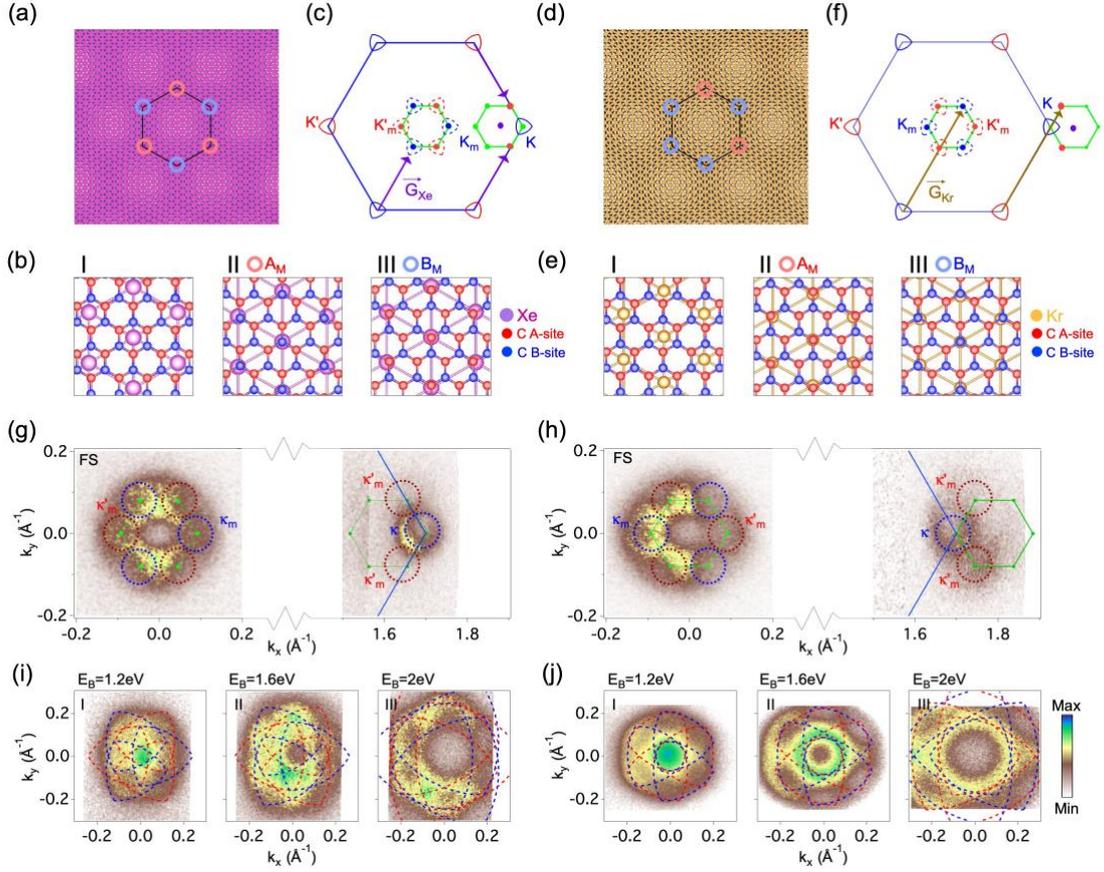

**Figure 1. High-order moiré patterns in mXe/G and mKr/G.** (a) Large-scale atom arrangements showing moiré pattern in mXe/G, with $a_{Xe} < \sqrt{3}a_{Gra}$. (b) Zoom-in areas at the center (I) and corners (II-III) of moiré pattern in mXe/G. (c) Schematic high-order moiré effect on band structure modulation in mXe/G. (d) Same as (a), but for mKr/G with $a_{Kr} > \sqrt{3}a_{Gra}$. (e) Same as (b), but for mKr/G. (f) Same as (c), but for mKr/G. (g, h) Fermi surface maps of mXe/G with $T_a$ = 67 K and mKr/G at $T_a$ = 37 K, respectively. (i, j) Constant energy intensity plots at different $E_B$ of mXe/G and mKr/G, respectively. The dashed lines of different colors indicate Dirac cone replicas from different valleys.

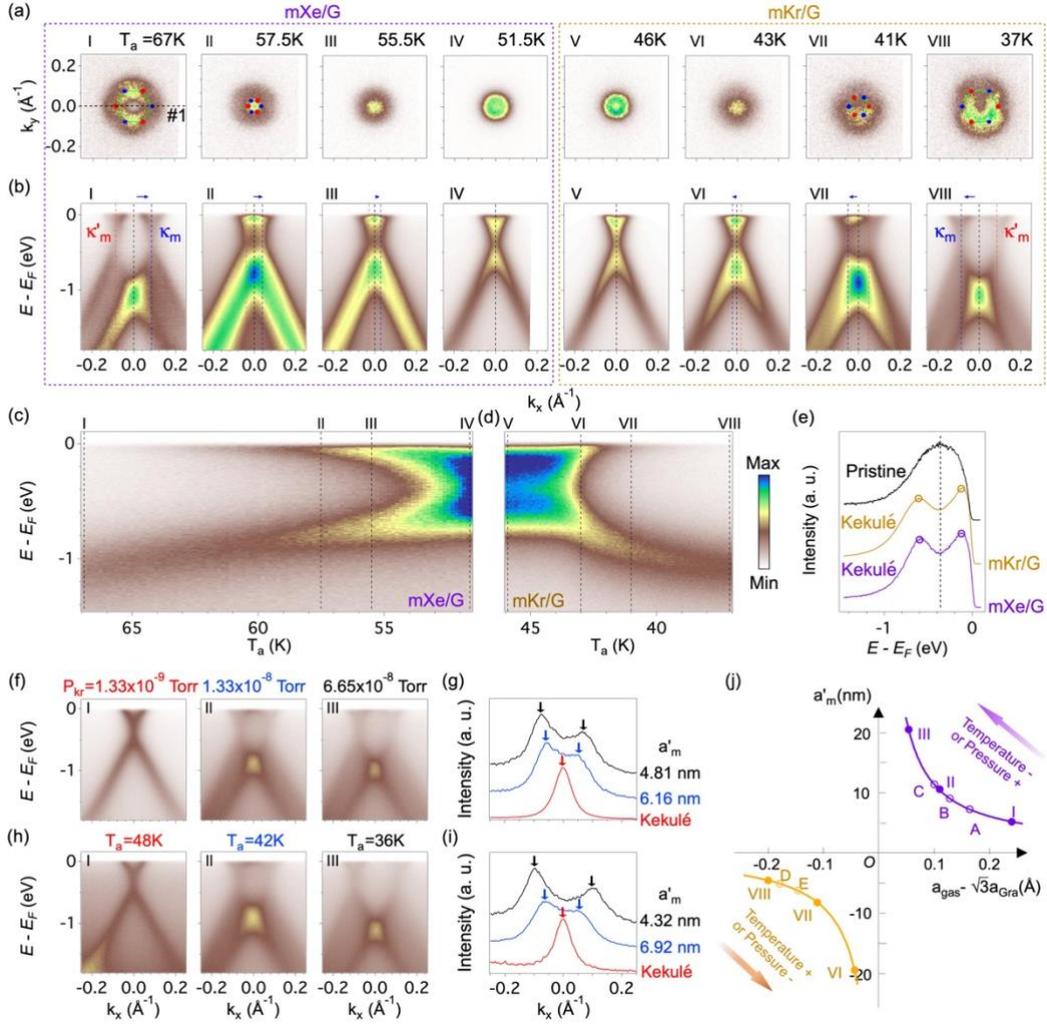

**Figure 2. *In situ* tunable moiré pattern. (a)** Fermi surface maps of mXe/G (I-IV) and mKr/G (IV-VII) at different $a'_m$. **(b)** Correspondingly band structures along the path #1 in (a)-I of mXe/G (I-IV) and mKr/G (IV-VII) at different $a'_m$. **(c, d)** The temperature dependent photoemission data at the Γ point of mXe/G and mKr/G. The dashed lines represent the photoemission data at the Γ point of mXe/G (I-IV) and mKr/G (IV-VII) in (b), respectively. **(e)** The energy distribution curves (EDCs) at the Γ point for the Kekulé distortion of mXe/G and mKr/G. The EDC at K point of pristine graphene is also plotted as black line. **(f)** Band structure along the Γ-K direction measured at T = 46 K, in Kr pressure $P_{Kr}$ = 1.33×10⁻⁹ torr (I), 1.33×10⁻⁸ torr (II) and 6.65×10⁻⁸ torr (III), respectively. (g) Corresponding momentum distribution curves (MDC) at Dirac point energy ($E_B$ = 0.36 eV). The derived moiré period $a'_m$ is also labeled. (h) Band structure along the Γ-K direction measured at T = 6 K, as mKr/G annealed at $T_a$ = 48 K (I), 42 K (II) and 36 K (III), respectively. (i) Similar to (g) but for different anneal temperature. (i) The moiré period as a function of $a_{Xe}$ (purple line) and $a_{Kr}$ (brown line). The data points I-III refer to (a)-I to (a)-III, and A-C refer to $P_{Xe}$ = 6.6×10⁻¹⁰ torr, 6.6×10⁻⁹ torr and 4.0×10⁻⁸ torr at $T_a$ = 60 K [26] in mXe/G. The data points VI-VIII refer to (a)-VI to (a)-VIII, and D-E refer to (f)-II to (f)-III.

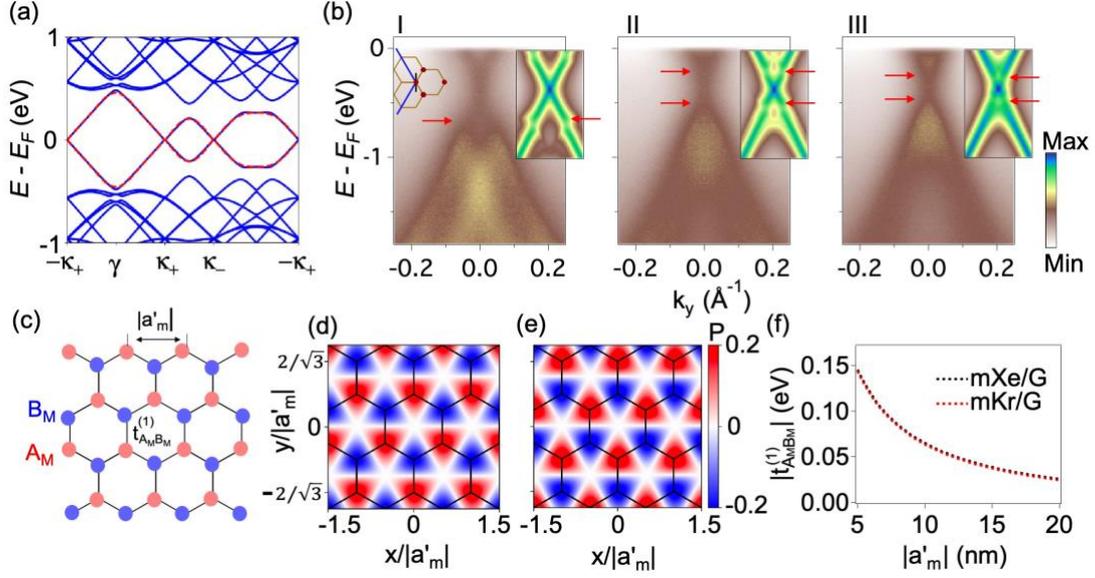

**Figure 3. Simulation of honeycomb lattice in mesoscale by high-order moiré patterns in mXe/G and mKr/G.** (a) Calculated moiré band structure of mKr/G along high-symmetry path in the moiré BZ. The red dashed lines show the tight-binding-model fit to the two middle bands. $|a'_m|$ is 5 nm in the calculation. The Dirac points are set to be at zero energy in this plot. (b) Experimentally determined band structures of mKr/G near the K point along the black line in top left corner of b-I at $a'_m$ = 5 nm, 10 nm and 20 nm, achieved by annealing at $T_a$ = 37 K, 42 K and 43 K, respectively. Corresponding theoretical simulation are appended in the insets for the direct comparison. (c) Schematic simulation of honeycomb lattice formed by $A_M$ and $B_M$ sites. $t^{(1)}_{A_M B_M}$ denotes the nearest-neighbour hopping. (d, e) The averaged sublattice polarization $P$ for the two middle bands in mXe/G and mKr/G, respectively. $|a'_m|$ is 5 nm in the calculation. (f) The calculated magnitude of the nearest-neighbour hopping parameter $t^{(1)}_{A_M B_M}$ as a function of $|a'_m|$ in mXe/G (purple dashed line) and mKr/G (brown dashed line).

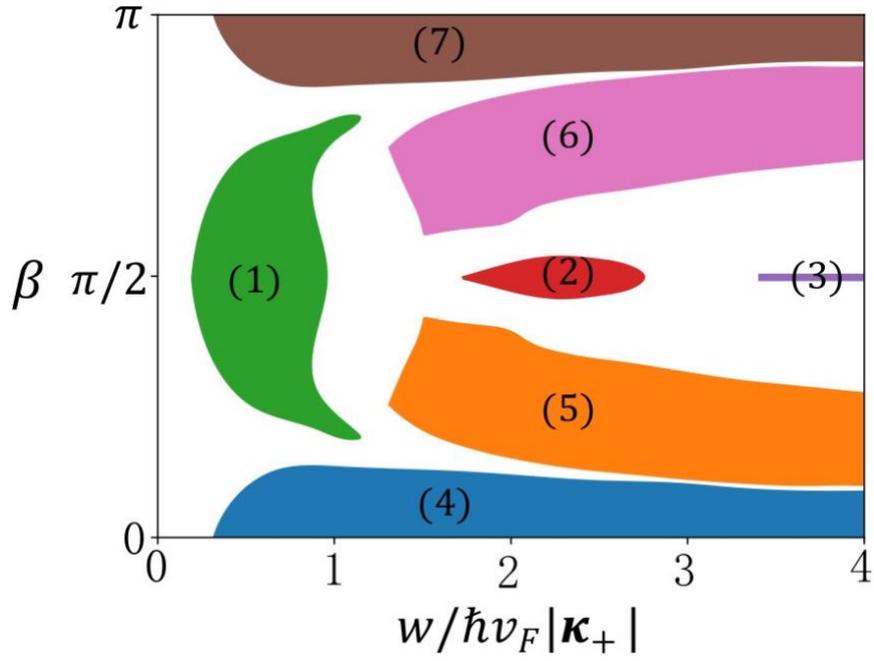

**Figure 4. Topological phase diagram of the two middle bands of moiré Hamiltonian $H$ as a function of parameters $w/\hbar v_F|\kappa_+|$ and $\beta$.** In the white regions, the middle bands are not separated by an overall energy gap from other bands. In the color shaded regions, the gapped phases (1)-(3) possess fragile topology, while the gapped phases (4)-(7) are topologically trivial and described by a tight-binding model defined on a honeycomb lattice.

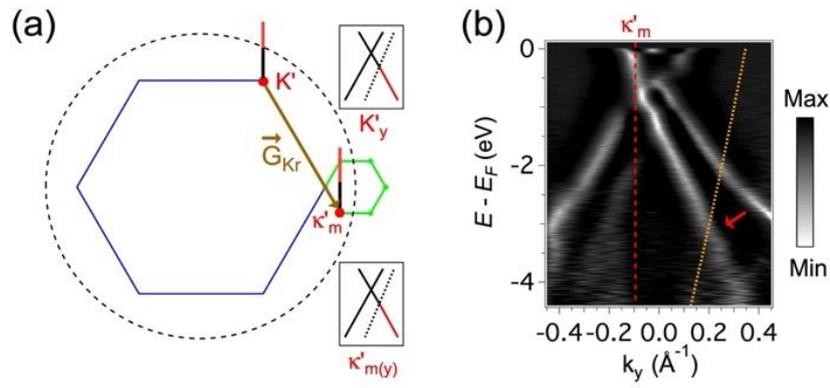

**Figure 5. The photoelectron diffraction scenario.** (a) Schematic diagram of final state diffraction picture in mKr/G. The black dashed line indicates the maximum measurement range of graphene in ARPES with a photon energy of 21.2eV. (b) The curvature of Dirac replicas $\kappa'_m$ along black solid line in a, the dashed line corresponds momentum cutoff limit.

# References

[1] Y. Cao, V. Fatemi, A. Demir, S. A. Fang, S. L. Tomarken, J. Y. Luo, J. D. Sanchez-Yamagishi, K. Watanabe, T. Taniguchi, E. Kaxiras, R. C. Ashoori, and P. Jarillo-Herrero, Correlated insulator behaviour at half-filling in magic-angle graphene superlattices. Nature **556,** 80 (2018).

[2] Y. Cao, V. Fatemi, S. A. Fang, K. J. Watanabe, T. Taniguchi, E. Kaxiras, and P. Jarillo-Herrero, Unconventional superconductivity in magic-angle graphene superlattices. Nature **556,** 43 (2018).

[3] G. R. Chen, L. L. Jiang, S. Wu, B. Lyu, H. Y. Li, B. L. Chittari, K. Watanabe, T. Taniguchi, Z. W. Shi, J. Jung, Y. B. Zhang, and F. Wang, Evidence of a gate-tunable Mott insulator in trilayer graphene moiré superlattice. Nat. Phys. **15,** 237 (2019).

[4] G. R. Chen, A. L. Sharpe, P. Gallagher, I. T. Rosen, E. J. Fox, L. L. Jiang, B. Lyu, H. Y. Li, K. Watanabe, T taniguchi, J. Jung, Z. W. Shi, D. Goldhaber-Gordon, Y. B. Zhang, and F. Wang, Signatures of tunable superconductivity in a trilayer graphene moiré superlattice. Nature **572,** 215 (2019).

[5] X. B. Lu, P. Stepanov, W. Yang, M. Xie, M. A. Aamir, I. Das, C. Urgell, K. Watanabe, T. Taniguchi, G. Y. Zhang, A. Bachtold, A. H. MacDonald, and D. K. Efetov, Superconductors, orbital magnets and correlated states in magic angle bilayer graphene. Nature **574,** 653 (2019).

[6] P. Stepanov, I. Das, X. B. Lu, A. Fahimniya, K. Watanabe, T. Taniguchi, F. H. L. Koppens, J. Lischner, L. Levitov, and D. K. Efetov, Untying the insulating and superconducting orders in magic-angle graphene. Nature **583,** 375 (2020).

[7] F. C. Wu, T. Lovorn, E. Tutuc, and A. H. Macdonald, Hubbard model physics in transition metal dichalcogenide moiré bands. Phys. Rev. Lett. **121,** 026402 (2018).

[8] M. H. Naik, and M. Jain, Ultraflatbands and shear solitons in moiré patterns of twisted bilayer transition metal dichalcogenides. Phys. Rev. Lett. **121,** 266401 (2018).

[9] D. M. Kennes, M. Claassen, L. Xian, A. Georges, A. J. Millis, J. Hone, C. R. Dean, D. N. Basov, A. N. Pasupathy, and A. Rubio, Moiré heterostructures as a condensed-matter quantum simulator. Nat. Phys. **17,** 155 (2021).

[10] H. C. Po, L. J. Zhou, A. Vishwanath, and T. Senthil, Origin of Mott insulating behavior and superconductivity in twisted bilayer graphene. Phys. Rev. X **8,** 031089 (2018).

[11] Z. D. Song, Z. J. Wang, W. J. Shi, G. Li, C. Fang, and B. A. Bernevig, All magic angles in twist bilayer graphene are topological. Phys. Rev. Lett. **123,** 036401 (2019).

[12] Z. D. Song, and B. A. Bernevig, Magic-angle twisted bilayer graphene as a topological heavy fermion problem. Phys. Rev. Lett. **129,** 047601 (2022).

[13] F. C. Wu, T. Lovorn, E. Tutuc, I. Martin, and A. H. MacDonald, Topological insulators in twisted transition metal dichalcogenide homobilayers. Phys. Rev. Lett. **122,** 086402 (2019).

[14] M. Angeli and A. H. Macdonald, Γ-valley transition-metal-dichalcogenide moiré bands. PNAS **10,** 118 (2021).

[15] Y. H. Tang, L. Z. Li, T. X. Li, Y. Xu, S. Liu, K. Barmak, K. Watanabe, T. Taniguchi, A. H. Macdonald, J. Shan and K. F. Mak, Simulation of Hubbard model physics in $WSe_2/WS_2$ moiré superlattices. Nature **579**, 353 (2020).

[16] E. C. Regan, D. Q. Wang, C. H. Jin, M. B. Utama, B. N. Gao, X. Wei, S. H. Zhao, W. Y. Zhao, Z. C. Zhang, K. Yumigeta, M. Blei, J. D. Calström, K. J. Watanabe, T. Taniguchi, S. Tongay, M. Crommie, A. Zettl, and F. Wang, Mott and generalized Wigner crystal states in $WSe_2/WS_2$ moiré superlattices. Nature **579**, 359 (2020).

[17] Y. Xu, S. Liu, D. A. Rhode, K. Watanabe. T. Taniguchi, J. Hone, V. Elser, K. F. Mak, and J. Shan, Correlated insulating states at fractional fillings of moiré superlattices. Nature **587**, 214 (2020).

[18] H. Y. Li, S. W. Li, E. C. Regan, D. Q. Wang, W. Y. Zhao, S. Kahn, K. Yumigeta, M. Blei, T. Taniguchi, K. Watanabe, S. Tongay, A. Zettl, M. F. Crommie, and F. Wang, Imaging two-dimensional generalized Wigner crystals. Nature **597**, 650 (2021).

[19] T. X. Li, S. W. Jiang, B. Shen, Y. Zhang, L. Z. Li, Z. Tao, T. Devakul, K. Watanabe, T. Taniguchi, L. Fu, J. Shan, and K. F. Mak, Quantum anomalous Hall effect from intertwined moiré bands. Nature **600**, 641 (2021).

[20] N. R. Wilson, K. V. Nguyen, K. Seyler, P. Rivera, A. J. Marsden, Z. L. Laker, G. C. Constantinescu, V. Kandyba, A. Barinov, N. D. M. Hine, X. D. Xu, and D.H. Cobden, Determination of band offsets, hybridization, and exciton binding in 2D semiconductor heterostructures. Sci. Adv. **3**, 1601832 (2017).

[21] S. Lisi, X. B. Lu, T. Benschop, T. A. D. Jong, P. Stepanov, J. R. Duran, F. Margot, I. Cucchi, E. Cappelli, A. Hunter, A. Tamai, V. Kandyba, A, Giamipetri, A. Barinov, J. Jobst, V. Stalman, M. Leeuwenhoek, K. Watanabe, T. Taniguchi, L. Rademaker, S. J. Molen, M. P. Allan, D. K. Efetov, and F. Baumberger, Observation of flat bands in twisted bilayer graphene. Nat. Phys. **17**, 189 (2021).

[22] M. L. B. Utama, R. J. Koch, K. Lee, N. Leconte, H. Y. Li, S. H. Zhao. L. Jiang, J. Y. Zhu, K. Watanabe, T. Taniguchi, P. D. Ashby, A. W. Bargioni, A. Zettl, C. Jozwiak, J. Jung, E. Rotenberg, A. Bostwick, and F. Wang, Visualization of the flat electronic band in twisted bilayer graphene near the magic angle twist. Nat. Phys. **17**, 184 (2021).

[23] Y. W. Li, S. H. Zhang, F. Q. Chen, L. Y. Wei, Z. L. Zhang, H. B. Xiao, H. Gao, M. Y. Chen, S. J. Liang, D. Pei, L. X. Xu, K. Watanabe, T. Taniguchi, L. X. Yang, F. Miao, J. P. Liu, B. Cheng, M. X. Wang, Y. L .Chen, and Z. K. Liu, Observation of Coexisting Dirac Bands and Moiré Flat Bands in Magic-Angle Twisted Trilayer Graphene. Adv. Mater. **10**, 1002 (2022).

[24] D. Pei, B. B. Wang, Z.S. Zhou, Z. H. He, L. H. An, S. M. He, C. Chen, Y. W. Li, L. Y. Wei, A. J. Liang, J. Avila, P. Dudin, V. Kandyba, A. Giampietri, M. Cattelan, A. Barinov, and Z. K. Liu, Observation of Γ-Valley moiré bands and emergent hexagonal lattice in twisted transition metal dichalcogenides. Phys. Rev. X **12,** 021065 (2022).

[25] C. L. Wu, Q. Wan, C. Peng, S. K. Mo, R. Z. Li, K. M. Zhao, Y. P. Guo, C. D. Zhang, and N. Xu, Observation of high-order moiré effect and multiple Dirac fermions replicas in graphene-SiC heterostructure. Phys. Rev. B **104**, 235130 (2021).

[26] C. L. Wu, Q. Wan, C. Peng, S. K. Mo, R. Z. Li, K. M. Zhao, Y. P. Guo, S. J. Yuan, F. C. Wu, C. D. Zhang, and N. Xu, Tailoring dirac fermions by in-situ tunable high-order moiré pattern in graphene-monolayer xenon heterostructure. Phys. Rev. Lett. **129**, 176402 (2022).

[27] S. Im, H. Im, K. Kim, J.-E. Lee, J. Hwang, S.-K. Mo, C. Hwang, Modified Dirac Fermions in the Crystalline Xenon and Graphene Moiré Heterostructure. Adv. Phys. Res. **2**, 2200091 (2023).

[28] Y. W. Li, Q. Wan, N. Xu, Recent Advances in Moiré Superlattice Systems by Angle-Resolved Photoemission Spectroscopy. Adv. Mater. 2305175 (2023).

[29] Q. Y. Wang, W. H. Zhang, L. L. Wang, K. He, X. C. Ma, and Q. K. Xue, Large-scale uniform bilayer graphene prepared by vacuum graphitization of 6H-SiC (0001) substrates. J Phys.: Condens. Mater. **25**, 095002 (2013).

[30] See Supplemental Material for LEED patterns, moiré Hamiltonian, estimation of model parameters, bloch states, band symmetries for mXe/G and mKr/G, wannier states and tight-binding model constructions, and topological phase diagram, which includes Refs. [26, 37-38].

[31] Q. Lu, C. Le, X. Zhang, J. Cook, X. He, M. Zarenia, M. Vaninger, P. F. Miceli, D. J. Singh, C. Liu, H. Qin, T.-C. Chiang, C.-K. Chiu, G. Vignale, and G. Bian, Dirac fermion cloning, moiré flat bands, and magic lattice constants in epitaxial monolayer graphene, Adv. Mater. 34, 2200625 (2022).

[32] V. Crépel, A. Dunbrack, D. Guerci, J. Bonini, and J. Cano, Chiral model of twisted bilayer graphene realized in a monolayer, arXiv:2305.14423 (2023).

[33] S. Raghu, X. L. Qi, C. Honerkamp, and S. C. Zhang, Topological Mott insulators. Phys. Rev. Lett. **100**, 156401 (2008).

[34] T. Grover, and T. Senthil, Topological spin hall states, charged skyrmions, and superconductivity in two dimensions. Phys. Rev. Lett. **100**, 156804 (2008).

[35] Z. Y. Meng, T. C. Lang, S. Wessel, F. F. Assaad, and A. Muramatsu, Quantum spin liquid emerging in two-dimensional correlated Dirac fermions. Nature **464**, 847 (2010).

[36] C. M. Polley, L. I. Johansson, H. Fedderwitz, T. Balasubramanian, M. Leandersson, J. Adell, R. Yakimova, C. Jacobi, Origin of the π-band replicas in the electronic structure of graphene grown on 4H-SiC(0001). Phys. Rev. B **99**, 115404 (2019).

[37] X-J. Luo, M. Wang, and F. Wu, Symmetric Wannier states and tight-binding model for quantum spin Hall bands in AB-stacked $MoTe_2/WSe_2$, Phys. Rev. B **107**, 235127 (2023).

[38] B. Bradlyn, L. Elcoro, J. Cano, M. G. Vergniory, Z. Wang, *C. Felser*, M. I. Aroyo, and B. A. Bernevig, Topological quantum chemistry, Nature **547**, 298 (2017).